\documentclass[useAMS,usenatbib]{mn2e}
\usepackage{psfig}
\usepackage{epsfig}
\usepackage{lscape}
\usepackage{rotating}
\usepackage{amsmath}
\usepackage{graphicx,color}
\usepackage{subfigure}
\usepackage{multicol}
\usepackage{multirow}

\title[A survey of nulling pulsars using the Giant Meterwave Radio Telescope]{A survey of 
nulling pulsars using the Giant Meterwave Radio Telescope}

\author[Vishal Gajjar et al.]
{Vishal Gajjar$^{1}$\thanks{e-mail:gajjar@ncra.tifr.res.in}, 
B. C. Joshi$^{1}$, M. Kramer$^{2}$\\
$^{1}$National Centre for Radio Astrophysics, Post Bag 3, Ganeshkhind, Pune 411 007, India \\
$^{2}$MPI fuer Radioastronomie, Auf dem Huegel 69, 53121 Bonn, Germany}

\date{\today}

\pagerange{\pageref{firstpage}--\pageref{lastpage}} \pubyear{2011}

\begin{document}

\maketitle

\label{firstpage}

\begin{abstract}
Several pulsars show sudden cessation of pulsed emission, 
which is known as pulsar nulling. In this paper, the  
nulling behaviour of 15 pulsars is presented. The nulling 
fractions of these pulsars, along with the degree of reduction 
in the pulse energy during the null phase, are reported for 
these pulsars. A  quasi-periodic null-burst pattern is 
reported for PSR J1738$-$2330. 
The distributions of lengths of the null and the burst phases as well as the typical 
nulling time scales are estimated for eight strong pulsars. 
The nulling pattern of four pulsars with similar 
nulling fraction are found to be different from each other, 
suggesting that the fraction of null pulses 
does not quantify the nulling behaviour of a pulsar in full detail. 
Analysis of these distributions also indicate that while the null and the burst pulses 
occur in groups, the underlying distribution of the interval between a  
transition from the null to the burst phase and vice verse appears to be 
similar to that of a stochastic Poisson point process.
\end{abstract}


\section{Introduction}
\label{intro}
The abrupt cessation of pulsed radio emission for 
several pulse periods, exhibited by some pulsars, 
has remained unexplained despite the discovery 
of this phenomenon in many radio pulsars. This 
phenomenon, called pulse nulling, was first 
discovered in four pulsars in 1970. 
Subsequent studies have revealed pulse nulling 
in about 100 pulsars to date (Backer 1970; 
Hesse \& Wielebinski 1974; Ritchings 1976; 
Biggs 1992; Vivekanand 1995; Wang et al. 2007). 
The degree and form of pulse nulling varies from 
one pulsar to another. On one hand, there are 
pulsars such as PSR B0826$-$34 (Durdin et al. 1979), 
which null most of the time, and PSR B1931+24 
(Kramer et al. 2006), which exhibits no radio emission for 
2 $-$ 4 weeks. In contrast, pulsars such as PSR B0809+74 
show a small degree of nulling (Lyne \& Ashworth 1983). 
Pulse nulling is frequent in pulsars such as PSR B1112+50, while it is very 
sporadic in PSR B1642$-$03 (Ritchings 1976).

The fraction of pulses with no detectable 
emission is known as the nulling fraction (NF) 
and is a measure of the degree of nulling in 
a pulsar. However, NF does not specify 
the duration of individual nulls, nor does it 
specify how the nulls are spaced in time. 
Although some attempts of characterizing 
patterns in pulse nulling were made in the 
previous studies (Backer 1970; Ritchings 1976;
Janssen \& van Leeuwen 2004; Kloumann \& Rankin 2010), 
not many pulsars have been studied for 
systematic patterns in nulling, partly because 
these require sensitive and long observations. 

Recent discoveries suggest that nulling pulsars 
with similar NF may have different null durations. 
These include intermittent pulsars, 
such as PSR B1931+24 (Kramer et al. 2006) and PSR J1832+0029  
(Lyne 2009), and the rotating radio transients (RRATs), 
which show no pulsed emission between single burst of emission 
(McLaughlin et al. 2006). These pulsars also 
show extreme degree of nulling similar to few classical nulling pulsars. 
PSR B1931+24 exhibits radio pulsations for 5 to 10 days 
followed by an absence of pulsations for 25 to 35 days 
(Kramer et al. 2006). If the cessation of radio emission in this 
pulsar is interpreted as a null, it has a NF of about $\sim$75 percent 
similar to PSR J1502$-$5653 (Wang et al. 2007). Yet the latter shows nulls 
with a typical duration of few tens of seconds in contrast 
to a much longer duration for PSR B1931+24. A similar 
conclusion can be drawn by comparing RRATs with classical high 
NF pulsars. While this leads to the expectation that pulsars 
with similar NF may have different nulling timescales, 
no systematic study of this aspect of nulling is available to 
the best of our knowledge. In this paper, a modest attempt to 
investigate this is initiated.

Pulse nulling was usually believed to be a random phenomenon  
(Ritchings 1976; Biggs 1992). However, recent studies indicate a 
non-random nulling behaviour for a  few classical nulling pulsars 
(Redman and Rankin 2009; Kloumann and Rankin 2010). 
Redman and Rankin (2009) also report random nulling behaviour 
for at least 4 out of 18 pulsars in their sample. Therefore, 
it is not clear if non-randomness in the sense defined in 
Redman and Rankin (2009) is seen in most nulling pulsars and 
such a study needs to be extended to more nulling pulsars. 
This issue is investigated in this paper with a distinct set of 
nulling pulsars.

In this paper, we present observations of 15 pulsars, 
carried out using the Giant Meterwave Radio Telescope (GMRT) at 
325 and 610 MHz. Among these, five were discovered in the Parkes multibeam 
pulsar survey (PKSMB; Manchester et al. 2001; Morris et al. 2002; 
Kramer et al. 2003; Lorimer et al. 2006), which have no previously reported 
nulling behaviour. Rest of the sample consists of well 
known strong nulling pulsars. 
The observations and analysis techniques 
are described in Section \ref{sec2}. In Section \ref{sec3}, 
interesting features of the nulling behaviour for a few pulsars 
are discussed along with estimates of their NF and the reduction 
in the pulsed energy during the null phase. 
A comparison of the null length and burst 
length distributions for pulsars, which have similar NF, 
is presented in Section \ref{sec4} and a discussion 
on the randomness of nulls is  
presented in Section \ref{sec5}. The expected time-scales 
for the null and burst durations are presented in Section 
\ref{sec5a}. 
Finally, 
the implications of these results are discussed in Section 
\ref{sec6} and the conclusions are presented in Section \ref{sec7}. 

%%%%%%%%%%%%%%%%%%%%%%%%%%%%%%% Profiles  %%%%%%%%%%%%%%%%%%%%%%%%%%
\begin{figure}
 \centering
\psfig{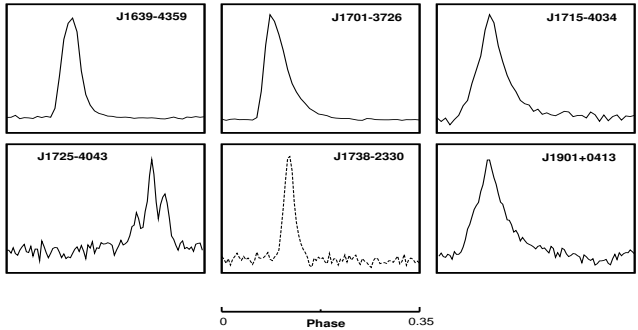} 
\caption{Integrated pulse profiles of 
six pulsars (five of which were discovered in PKSMB), 
with no previously reported low frequency study, at 325/610 MHz. 
The profile for PSR J1738$-$2330 (dashed line) was obtained from 
data at 325 MHz, while the other profiles (continuous line) 
were obtained from data at 610 MHz. All the profiles are plotted 
over equal phase interval (indicated at the bottom of the figure) 
 for comparison. }
\label{profilesfig}
\end{figure}
%%%%%%%%%%%%%%%%%%%%%%%%%%%%%%%%%%%%%%%%%%%%%%%%%%%%%%%%%%%%%%%%%%%%%%%%%%%%%

\section{Observations and Results}
\label{sec2}
This survey of a sample of nulling pulsars was conducted 
using the GMRT (Swarup et al. 1991). The observations were carried 
out at 325 and 610 MHz using a total time of 60 hrs 
from 2008 November 9 to 2009 August 21. For our 
observations, the GMRT correlator was used as a 
digital filterbank to obtain 256 spectral channels, each 
having a bandwidth of 62.5 KHz, across the 16 MHz bandpass 
for each polarization received from the 
30 antennas of GMRT. The digitized signals from typically 
14 to 20 GMRT antennas were added in phase (the GMRT was 
used as a phased array) forming a coherent 
sum of signals with the GMRT Array Combiner (GAC).  
Then, the summed signals were detected in each channel. The 
detected powers in each channel were then acquired into 
16-bit registers every 16 $\mu$s after summing the two 
polarizations in a digital backend and were accumulated 
before being written to an output buffer to reduce the 
data volume. The data were then acquired using a data 
acquisition card and recorded to a  tape for off-line 
processing. The effective sampling time for this 
configuration, used for most of the survey, was 1 msec. 

In the off-line processing, the data were first dedispersed 
and then folded to typically 256 phase bins across the 
topocentric pulse period using a publicly available package 
SIGPROC\footnote{http://sigproc.sourceforge.net} to obtain an 
integrated profile for each pulsar. 
Our sample included six pulsars (among which five were discovered in PKSMB pulsar survey)
with no previously reported low frequency studies. 
The integrated profiles for these six 
pulsars are shown in Figure \ref{profilesfig}. These profiles 
at low radio frequencies are being reported for the first time 
to the best of our knowledge. Two of these pulsars, 
PSRs J1701$-$3726 and J1901+0413, exhibit marked scattering 
tail in their profiles at 610 MHz, while scattering tails 
are also apparent in two other pulsars, PSRs J1639$-$4359 
and J1715$-$4034, although not as prominent as in the case of the former 
two pulsars. 

\begin{figure}
 \centering
\psfig{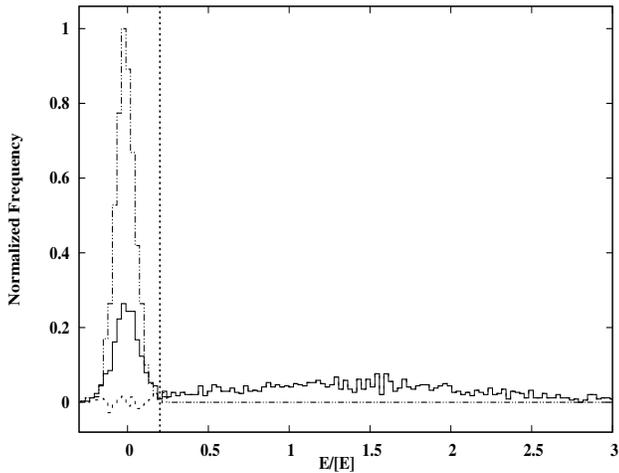}
% Histrogram2_forpaper.pdf: 433x618 pixel, 72dpi, 15.28x21.80 cm, bb=0 0 433 618
 
\caption{Normalized histogram of on-pulse (solid line) and off-pulse (dot-dot-dashed line) 
energies, normalized as explained in the text, for PSR B2319+60. The residuals after 
the subtraction of a scaled version of off-pulse distribution from the on-pulse distribution 
are shown with dashed line. The dotted vertical line is visually selected threshold at the 
point where null and burst pulse distributions cross each other in on-pulse histogram.}
\label{histogram}
\end{figure}

The dedispersed data were then folded to typically 256 phase 
bins across the topocentric pulse period for each period to 
form a single pulse sequence. For few pulsars with low 
intensity single pulses, a fixed number, N, of successive pulses 
(indicated within parentheses in Column 10 of Table \ref{Results}) 
in the single pulse data were averaged to form average pulse for 
every block of N single pulses (hereafter called subintegration),  
to obtain sufficient signal to noise ratio 
(i.e. S/N $\geq$ 5 for each subintegration). 
Pulses affected by radio frequency interference (RFI), were 
removed before the NF analysis. The NF for each pulsar was then obtained 
using a procedure similar to that used for detecting pulse nulling in 
single pulse sequences (Ritchings 1976; Vivekanand 1995). 
A baseline, estimated using bins away from the on-pulse bins, 
was subtracted from the data for each pulse. The 
total energy for each pulse in the on-pulse window and the off-pulse window 
(both with equal number of bins) was calculated from the baseline 
subtracted data. The energies thus obtained as a function of 
period number form an on-pulse energy (ONPE) sequence and 
off-pulse energy (OFPE) sequence. Next, the ONPE and OFPE 
were scaled by the average on-pulse energy, calculated for 
every block of 200 periods from ONPE, to compensate for 
variations due to inter-stellar scintillations. 
The normalized ONPE and OFPE were then binned to 
an appropriate number of energy bins depending upon 
the available S/N for a given pulsar. The fraction 
of total pulses per energy bin was plotted as a 
histogram as shown in Figure \ref{histogram}. 
An excess at zero energy in the on-pulse energy 
distribution indicates the fraction of nulled 
pulses or NF of the pulsar. This can be estimated 
by removing a scaled version of off-pulse 
energy distribution (modeled as a Gaussian) 
at zero energy from the on-pulse distribution. 
As the null and burst pulse distributions are not 
well separated for most of our pulsars, 
such a fit of the off-pulse energy distribution, 
modeled as a Gaussian, will have a corresponding 
error while scaling the on-pulse energy distribution near the  
zero pulse energy. This is reflected in the quoted errors on the NFs given in   
the parentheses in Column 6 of Table \ref{Results}. 

Three pulsars in our sample were weak. Hence, consecutive pulses were 
added (indicated inside parentheses of Column 10 in Table \ref{Results}) 
to improve the S/N ratio. NF for these three pulsars, 
obtained using the above mentioned method on data averaged 
over given number of consecutive periods (subintegration), 
will be an under estimate. 
Hence, only the lower limits on the NFs for these pulsars are presented 
in Table \ref{Results}. 

Only upper limits on the NFs could be estimated for three pulsars 
in the sample. Two of these were weak pulsars with significantly low  
nulling. Estimation of the lower limits on the NFs (as explained in 
the previous paragraph) was not possible as averaging over 
consecutive periods resulted in detection of emission for every 
subintegration giving a NF of zero. However, some of 
the single pulses in a subintegration may well have been nulled pulses. 
To estimate an upper limit on the fraction of such pulses 
for these two pulsars, we arranged all the single pulses in the 
ascending order of their on-pulse energy. A threshold was moved 
from high to low on-pulse energy end till the pulses below the 
threshold did not show a collective profile with a pulse at 
 significance greater 
than  3 times the root mean square deviation (RMS). 
All the pulses below such threshold were tagged as null pulses. 
The fraction of these pulses give an estimate for the upper limit on 
the NF and this limit is presented for these two pulsars 
in Column 6 of Table \ref{Results}. The reason for the upper limit on the 
NF for the remaining pulsar, PSR J1725$-$4043,  is explained in 
Section \ref{sec3}. 

%%%%%%%%%%%%%%%%%%%%%%%%%%%%%%%%%%%%%%%%   Histogram %%%%%%%%%%%%%%%%%%%%%%%%%%%%%%%%%%%%%%%%%%%%%%%%%%
\normalsize
%%%%%%%%%%%%%%%%%%%%%%%%%%%%%%%%%%%%%%%%%%%%%%%%%%%%%%%%%%%%%%%%%%%%%%%%%%%%%%%%%%%%%%%%%%%%%%%%%%%%%%%%
The degree, $\eta$, by which the radio emission from a 
nulling pulsar declines during the nulls can be 
obtained by forming the average profiles of the 
pulsar for burst and null pulses separately.  
A threshold energy, decided by examining the 
on-pulse and off-pulse energy distributions of 
the pulsar, was used to separate burst and 
null pulses as shown in Figure \ref{histogram}. 
If the distribution of burst and null 
pulses are not well separated, such a threshold will 
cause few low energy burst pulses to be tagged 
as null pulses and vice-verse. To identify these 
low energy burst pulses among null pulses, 
we arranged all null pulses in ascending order of 
their on-pulse energy. Pulses were removed manually 
from the high energy end till the null pulse profile, averaged 
over the remaining null pulses,  
did not show any profile component with more than 3 times the 
RMS. These pulses, from the high energy end, were tagged as 
burst pulses. 
Similarly, burst pulses were also arranged in ascending order of their on-pulse 
energy. Pulses from the lower end of the on-pulse energy, 
which did not form an average pulse profile with a significant pulse 
($\geq$ 3 times the RMS), 
were also removed and tagged as null pulses. 
We calculated $\eta$ for a pulsar in a manner similar 
to that described in Vivekanand \& Joshi (1997). First,  
the total energy in the on-pulse 
bins for the burst pulse profile was obtained. Then,  
an upper limit, estimated as three times the RMS, 
was obtained on the detectable emission in 
the null pulse profile. 
The ratio between these two quantities 
is defined as $\eta$ (Equation \ref{R}).

\begin{equation}
\eta ~ = ~ \frac {\sum\limits^{N}_{i=1} P_{bpulse} (i) } {3 \times RMS_{npulse} }
\label{R}
\end{equation}
where \\
P$_{bpulse}$ = Intensity in i$^{th}$ bin for the on-pulse window of the burst pulse profile \\
RMS$_{npulse}$ = RMS estimated over the on-pulse window of the null pulse profile \\
N = Number of bins in the on-pulse window \\
 
The errors given in the parentheses
in Column 8 were estimated as 3 times the off-pulse RMS in the  
burst pulse profile. We are reporting $\eta$ for 11 pulsars in our 
sample for the first time. 

The results of our analysis for the sample of pulsars 
in this study are presented in Table \ref{Results}. 
The Table presents some of the basic parameters for the observed pulsars 
(i.e. Column 3 : Period, Column 4 : DM and Column 5 : Flux density at 
1420 MHz). It also presents NFs with the corresponding error in the 
parentheses (i.e. Column 6) for all the pulsars of our sample. 
The previously reported NFs (i.e. Column 7) for some pulsars are also listed 
for comparison and our results are consistent with those 
reported earlier for these pulsars. NFs are being reported 
for the first time in five pulsars, which were discovered in PKSMB survey. 
For 11 pulsars in our sample, estimates for $\eta$ are reported for the first 
time (i.e. Column 8). 

To investigate the time scales of 
the null  and the burst phase (i.e. normal emission), 
the single pulse sequences for eight pulsars were examined. 
These pulsars show high S/N (i.e. $\geq$ 5) for single burst pulses.  
Hence, null and burst pulses can be separated easily using the above mentioned 
method. The null lengths and burst lengths as well as the total 
number of uninterrupted sequences of null 
and burst phases were obtained from this examination of data.  
The null length histogram (NLH) and burst length 
histogram (BLH) for these eight pulsars, 
are shown in Figure \ref{nbhistall}.

%%%%%%%%%%%%%%%%%%%%%%%%  J1725 J1715 J1738 sp %%%%%%%%%%%%%%%%%%%%%%%%%%%%%%%%%%%%%%
\begin{figure*}
 \centering
 \psfig{figure=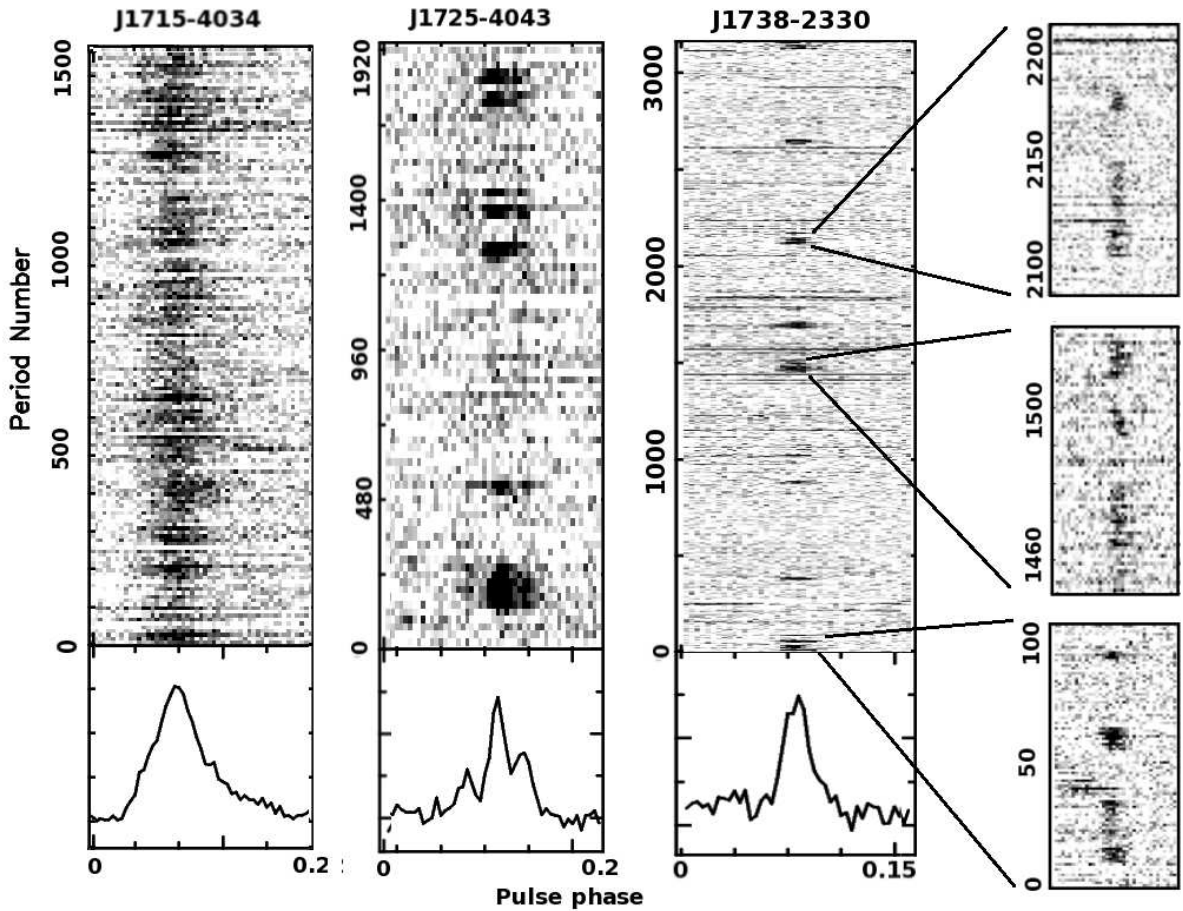,width=6.5in,height=5.0in,angle=0}
\caption{The modulation of 
pulsar energy in single pulses for three pulsars. Successive 10, 24 and 5 periods    
were averaged as a subintegration. The resulting subintegrations are plotted as gray-scale intensities for PSRs J1715$-$4034, J1725$-$4043 and 
J1738$-$2330 respectively for the entire stretch of observed data. 
Bottom panel of each plot shows the integrated profiles for the respective pulsars. 
The insets, where individual periods without any averaging are plotted, show 3 burst pulse bunches for PSR J1738$-$2330.}
\label{j1725sp}
\label{j1738sp}
\end{figure*}
%%%%%%%%%%%%%%%%%%%%%%%%%%%%%%%%%%%%%%%%%%%%%%%%%%%%%%%%%%%%%%%%%%%%%%%%%%%%%%%%%%%%%%%%%%%%%%%%%%%%%%%

%%%%%%%%%%%%%%%%%%%%%%%%%%%%%%%%%%%%%%%%%%%% Table of Results %%%%%%%%%%%%%%%%%%%%%%%%%%%%%%%%%
\begin{center}
\begin{table*}
{\small
\caption{Parameters for the pulsars, observed in this survey, 
along with the obtained nulling fraction (NF) and reduction 
in the pulse energy during the null phase ($\eta$). 
Columns give pulsar name at 2000 and 1950 epochs, period 
(P)$^{a}$, dispersion measure (DM)$^{a}$, flux density at 
1400 MHz (S1400)$^{a}$, NF (as defined in text) obtained in this study, NF reported 
previously, estimate of $\eta$ (as defined in text) obtained in this study, number of runs 
and the number of pulses used (N) along with the 
number of contiguous pulses integrated (given 
in parentheses) for the analysis. The error bars on obtained values of NF and $\eta$
are indicated after the estimates by the number in the round parentheses and represent 
3 times the standard deviation errors. The references for the previously 
reported NF in Column 7 are as follows: (1) Lyne \& Ashworth 1986 
(2) Biggs 1992 (3) Ritchings 1976 (4) Herfindal \& Rankin 2009 
(5) Wang et al. 2007}
\hfill{}
\label{Results}
\newline
\footnotesize
% \scriptsize
% \centering
\begin{tabular}[ht]{||c|c|c|c|c|c|c|c|c|c|}
\hline

J2000 	     & B1950 	 & Period  & DM          & S1400 &Obtained& Known&  $\eta$  & Number of  &  N (Sub-integration) \\
Name  	     & Name  	 &         &             &       & NF    & NF    &          & Runs       &                      \\  
 	     &		 & (s)	   & (pc/cm$^3$) & (mJy) & (\%)  & (\%)  &   $-$    & $-$        & $-$			\\ 
\hline
\hline
J0814+7429   &  B0809+74   & 1.292241 & 06.1  & 10.0 & 1.0(0.4) & 1.42(0.02) $^{[1]}$   & 172.0(0.5)   & 246 & 13766 (1)  \\
J0820$-$1350 &  B0818$-$13 & 1.238130 & 40.9  & 7.0  & 0.9(1.8) & 1.01(0.01) $^{[1]}$   & 4.2(0.2)     & 114  & 3341 (1)  \\
J0837$-$4135 &  B0835$-$41 & 0.751624 & 147.2 & 16.0 & 1.7(1.2)   & $\leq$1.2 $^{[2]}$    & 15.7(0.2)  & 148  & 3335 (1)  \\
J1115+5030   &  B1112+50   & 1.656439 & 9.2   & 3.0  & 64(6)     & 60(5) $^{[3]}$       & 44.7(0.2)  & 1270 & 2634 (1)  \\
J1639$-$4359 &    $-$      & 0.587559 & 258.9 & 0.92 & $\leq$0.1  & $-$	                  & $-$         & $-$ & 13034 (1)  \\
J1701$-$3726 &    $-$	   & 2.454609 & 303.4 & 2.9  & 19(6)     & $\geq$14 $^{[5]}$     & 6.4(0.2)   & $-$ & 2464 (1)  \\
J1715$-$4034 &    $-$      & 2.072153 & 254.0 & 1.60 & $\geq$6    & $-$                   & $-$         & $-$ & 1591 (16) \\
J1725$-$4043 &    $-$      & 1.465071 & 203.0 & 0.34 & $\leq$70   & $-$		          & $-$         & $-$ & 2481 (24) \\
J1738$-$2330 &	  $-$      & 1.978847 & 99.3  &	0.48 & $\geq$69   & $-$	                  & 5.3(0.3)   & $-$ & 2178 (5)  \\
J1901+0413   &    $-$      & 2.663080 & 352.0 & 1.10 & $\leq$6          & $-$                   & $-$         & $-$ & 2605 (1) \\
J2022+2854   &  B2020+28   & 0.343402 & 24.6  & 38   & 0.2(1.6)  & $\leq$ 3 $^{[3]}$     & 2.5(0.2)   & $-$ & 8039 (1)  \\
J2022+5154   &  B2021+51   & 0.529196 & 22.6  & 27.0 & 1.4(0.7)   & $\leq$5 $^{[3]}$      & 2.6(0.2)   & 24  & 1326 (1)  \\
J2037+1942   &  B2034+19   & 2.074377 & 36.0  & $-$  & $\geq$26   & 44(4) $^{[4]}$       & 6.4(0.1)   & 672 & 1618 (3)  \\
J2113+4644   &  B2111+46   & 1.014685 & 141.3 & 19.0 & 21(4)     & 12.5(2.5) $^{[3]}$   & 14.9(0.3)  & 290 & 6208 (1)  \\
J2321+6024   &  B2319+60   & 2.256488 & 94.6  & 12.0 & 29(1)      & 25(5) $^{[3]}$       & 115.8(0.4) & 450 & 1795 (1)  \\
\hline
\end{tabular}}
\hfill{}
\raggedright $^{a}$ ATNF Catalogue : http://www.atnf.csiro.au/research/pulsar/psrcat/
\end{table*}
\end{center}
%%%%%%%%%%%%%%%%%%%%%%%%%%%%%%%%%%%%%%%%%%%%%%%%%%%%%%%%%%%%%%%%%%%%%%%%%%%%%%%%%%%%%%%%%%%%%%%%%%%%%%%

\section{Nulling behaviour of individual pulsars}
\label{sec3}
\begin{figure}
 \centering
 \includegraphics[height=2.5in ,angle=-90,bb=14 14 520 736]{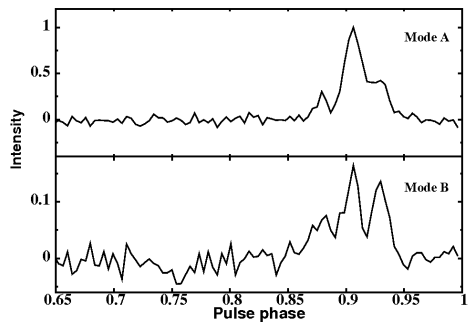}
\caption{Integrated pulse profiles for Mode A and Mode B of PSR J1725$-$4043. 
The ordinate is in arbitrary units and was obtained after scaling 
the two profiles by the peak intensity of the Mode A profile. 
The Mode B is around 10 times weaker than Mode A.}
\label{Both_mode}
\end{figure}

\begin{figure*}
 \centering
 \psfig{figure=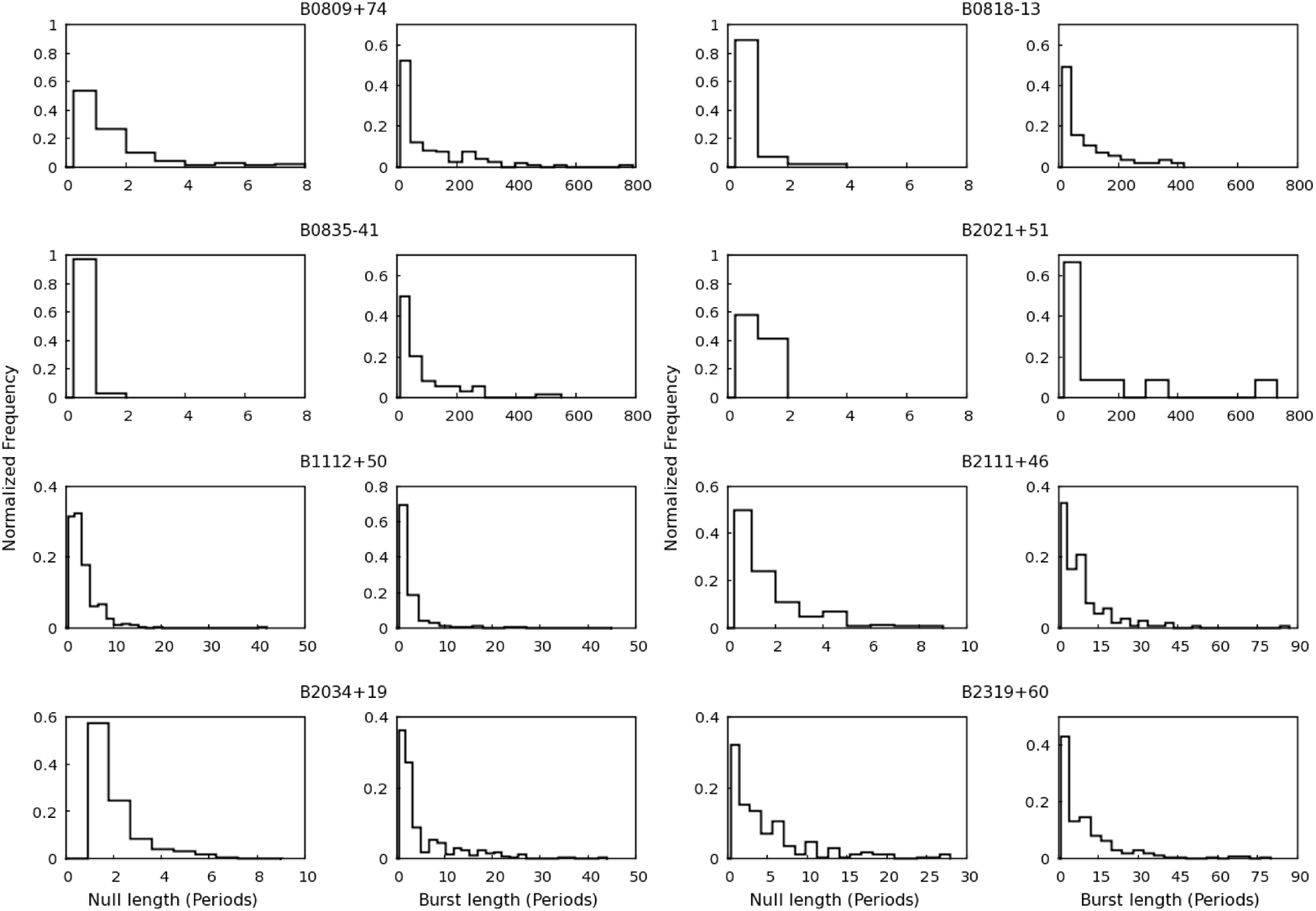,width=7.0in,height=5.5in}
\caption{Null length and burst length histograms of eight pulsars in the 
sample. The histogram in the left panel for each pulsar shows the distribution of observed null lengths, 
while that on the right panel shows the distribution of observed burst lengths.}
\label{nbhistall}
\end{figure*}

The salient features of the nulling behaviour of a few individual pulsars 
are discussed below
\newline
\newline
{\bf PSR J1715$-$4034 :}
This pulsar was discovered in PKSMB (Kramer et al. 2003).  
No nulling behaviour has been reported previously for this pulsar. 
Width of the pulses is significantly modulated as seen in Figure \ref{j1725sp}. 
There are two components in the integrated profile at 1420 MHz
(Kramer et al. 2003), which are difficult 
to identify at 610 MHz because of scatter-broadening. Independent modulation 
in these components can cause this apparent change in the 
pulse width. 
\newline
\newline
{\bf PSR J1725$-$4043 :}
This is also one of the pulsars discovered in PKSMB 
(Kramer et al. 2003) with no previously reported 
nulling behaviour. Its integrated profile at 610 MHz  
exhibits three narrow components - one strong central 
component with weak trailing and leading components.
Single pulse data were averaged over  
successive pulses to form average profiles for every block of 24 
single pulses (subintegrations). A gray-scale plot of these 
subintegrations as well as the integrated profile for 
this pulsar is shown in Figure \ref{j1725sp}. 
Visual inspection of the gray-scale plot shows emission 
bunches of 50 to 200 strong pulses, which correspond to the 
normal integrated profile (hereafter referred as Mode A - 
for example subintegrations 120 to 
264 in Figure \ref{j1725sp}). After adding all the 
subintegrations during the null phase, separated 
by visual inspection, a weak profile (hereafter referred as Mode B - for 
example subintegrations 264 to 480 in Figure \ref{j1725sp}), different from 
the normal integrated profile (Mode A), is obtained. 
The integrated profiles for the two modes are shown in Figure \ref{Both_mode}.  
Although the two profiles are 
similar in shape with distinct three components, Mode B profile 
shows relatively stronger trailing component (Figure \ref{Both_mode}). 
Hence, it appears that the pulsar shows sporadic 
emission with two distinct modes. 

Manual inspection of the single pulses, forming the null 
subintegrations,  reveals weak individual pulses  
among nulled pulses. 
Hence, it is difficult to identify null pulses as these could be 
low intensity Mode B pulses. Pulsar spends 30\% of time in Mode A 
emission. The remaining 70\% could be combination of null pulses 
and Mode B emission. Hence, only an upper limit on 
the NF for this pulsar is quoted in Table \ref{Results}. 
%%%%%%%%%%%%%%%%%%%%%%%%%%%%%%%%%% j1725 Modes %%%%%%%%%%%%%%%%%%%%%%%%%%%%%%%%%%%%%%%%%%%%%%
%%%%%%%%%%%%%%%%%%%%%%%%%%%%%%%%%%%%% NLH-BLH %%%%%%%%%%%%%%%%%%%%%%%%%%%%%%%%%%%%%%%%%
\newline
\newline
%%%%%%%%%%%%%%%%%%%%%%%%%%%%%%%%%%%%%%%%%%%%%%%%%%%%%%%%%%%%%%%%%%
{\bf PSR J1738$-$2330 :}
This is another pulsar discovered in the PKSMB 
(Lorimer et al. 2006). In our survey, it was observed at 325 MHz.
The pulsar seems to have quasi-periodic bursts, 
with an average duration of around 50 to 100 periods,  
interspersed with nulls of around 300 to 400 periods. 
This interesting single pulse behaviour in this pulsar is evident 
in Figure \ref{j1738sp}, where a gray scale plot,  
with 5 successive single pulses integrated, is shown.
We note that the bursts are quasi-periodic, and 
that a phase-resolved frequency spectrum (Figure 6) 
shows power at 0.0019 cycles/period and 0.0028 
cycles/period, which correspond to periodicities of 
approximately 525 and 350 periods, respectively. 
This 
quasi-periodic behaviour is similar to PSR B1931+24 but with 
much shorter time scale.

 \begin{figure}
 \centering
 \includegraphics[width=5cm,height=7cm,angle=90,bb=14 14 519 699]{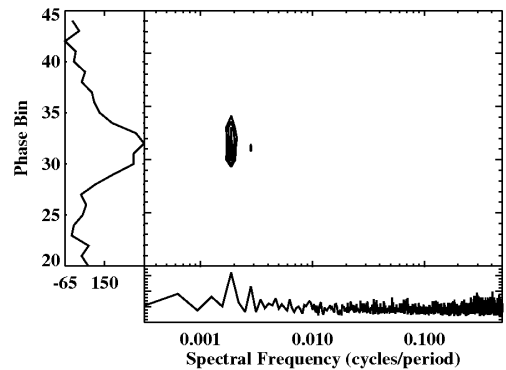}
 \caption{The phase-resolved spectra for PSR 
J1738$-$2330. The contours in the main panel show spectra for 
bins 20 to 44 across the on-pulse window. The left 
panel shows the integrated profile for these pulse 
phases, whereas  the bottom panel shows the averaged 
spectra for all the on-pulse phases (bins 28 to 38). 
The spectra in the bottom panel and in contours 
are shown with the abscissa plotted in a logarithmic scale upto the Nyquist frequency.}
\label{j1738lrf}
\end{figure}

Close examination of single pulses suggests that the bursts 
typically consist of a sequence of a 20 to 30 period 
burst followed by 2 short bursts of 5 to 10 pulses. 
These three bursts are separated by shorter nulls of 
about 10 to 25 pulses. The overall burst pattern is itself 
separated by around 400 pulse period nulls. Three examples 
of this burst pattern are indicated in Figure \ref{j1738sp}. 
However, our S/N was too low to confirm this with high 
significance. More sensitive and long observations in future may be 
useful to resolve burst pattern and periodicities. 
\newline
\newline
{\bf PSR B2020+28 :}
This is one of the well studied pulsars. The 
pulse profile shows two strong components, 
but this pulsar was classified as triple profile class
pulsar because of the prominent saddle region between the two 
components (Rankin et al. 1989). Interestingly, 
the emission in the individual profile components 
reduces significantly for a fraction of pulses, which is 
different for the two components. Our time resolution was not 
sufficient to identify the saddle region clearly so we 
were able to estimate NF for two components only. The 
estimated NF for the leading component is 3.5 $\pm$ 0.8 \% 
and $\eta$ is around 9.7.  The estimated NF for the trailing 
component is 9 $\pm$ 1 \% and $\eta$ is around 21, suggesting 
that at least the trailing component shows a nulling behaviour 
similar to a regular nulling pulsar. Both these individual NFs 
were significantly higher than the overall NF (Table \ref{Results}). 
\newline
\newline
{\bf PSR B2111+46 :}
This pulsar has a multicomponent profile, classified as a triple 
component profile class by Rankin et al. (1989). 
Like PSR B2020+28, the fraction of pulses for which emission 
is not detected varies from component to component.  
For the core component, NF was estimated to be of 26 $\pm$ 
2 \%, slightly higher than the NF of the entire pulse (Table 
\ref{Results}). The estimated $\eta$ value for this component is 
around 56. For the leading component, NF was found to be 
53 $\pm$ 5, almost double the NF for the entire 
pulse. NLH and BLH considering the emission in the entire 
pulse are reported here for the first time for this pulsar. 
NLH shows 50\% of nulls are single period nulls while other 
50\% are gradually distributed up to 10 periods.

\section{Comparing nulling behaviour}
\label{sec4}
In this study, NLH and BLH were obtained for 8 out of the 15 
pulsars observed. This allows a comparison of 
null lengths for these pulsars, particularly for 
PSRs B0809+74, B0818$-$13, B0835$-$41 and B2021+51, 
all of which have similar NF of around 1\%. 
Visual inspection of all single pulses for these 4 pulsars 
suggests that the pattern of arrangement of nulls within the 
burst differs for each of these pulsars and motivates a quantitative study. 
To investigate this, the NLHs and BLHs for these 
pulsars were examined. While the BLHs are similar 
for all the 4 pulsars, the NLHs show significant differences.
While PSRs B0835$-$41 and B2021+51 show only 
single and double period nulls, PSRs B0809+74 and B0818$-$13 
show nulls of longer durations as well. 

To quantify these differences, a two sample Kolmogorov-Smirnov 
(KS) test (Press et al. 2001) was carried out for the four pulsars mentioned above. 
KS-test is a non-parametric distribution free test, applicable to unbinned data. 
KS-test provides a statistic, D, which is the maximum difference between the two cumulative 
distribution functions (CDFs). To compare the null length distribution 
of two pulsars, we formed their CDFs from the observed unbinned null length 
sequences. We found D from their CDFs and estimated the rejection probability 
of null hypothesis, which assumes that the measured distributions are drawn from the same underlying 
distribution. The results of this test are given in Table 
\ref{K-S comparison} for each pair of the four pulsars. 
The null hypothesis is rejected for all pairs with 
high significance (except for the comparison of null lengths for 
PSRs B0818$-$13 and B0835$-$41, where the significance is marginally smaller).
Thus, the nulling patterns differ between each of the four pulsars 
even though they have the same NF. This kind of differences are not only seen in 
pulsars with small NF, but also in pulsars with larger NF. 
For example, PSRs B2319+60 and B2034+19 have NF $\sim$  
30\% but their NLH are different.  
A KS-test (result not shown in the table) once again 
rejects the null hypothesis with high significance. 
Likewise, NLH for PSRs B2111+46 also differ with PSRs B2319+60 and B2034+19. 
However, as these pulsars have slightly different 
NFs, it is difficult to draw a strong conclusion from 
these data.

\section{Do nulls occur randomly ?}
\label{sec5}
Previous studies (Redman and Rankin 2009; Kloumann and Rankin 2010)
indicated that nulling may not be random. To test the above 
premise, non-randomness tests were carried out on our data 
for 8 pulsars where it was possible to obtain NLH and BLH. 

If the null pulses of a pulsar are characterized by 
an independent identically distributed (iid) 
random variable, for which NF represents the proportion 
statistics, then one can Monte-Carlo simulate synthetic 
data sets using a random number generator (Press et al. 2001). 
We simulated around 10,000 random one-zero time series of the same   
length as that of the sequence of the observed pulses for each pulsar 
with a given NF. A distribution of null and burst length was 
derived from the synthetic data set. If the underlying 
distribution of observed null lengths does not differ from the 
the simulated distribution with high significance, then it can be concluded that 
the observed nulls are sampled from a  
distribution characterizing such an iid random variable, 
for which NF represents the proportion statistics. The above premise can be tested by carrying 
out a one sample KS-test (Press et al. 2001). As explained in the earlier section, 
KS-test provides D statistic, which is the maximum deviation between the two CDFs. 
 For our test, one CDF was obtained from a simulated null sequence 
while the other CDF was obtained from the observed null sequence. 
As usual, the test was carried out on the unbinned data. 
The D statistic from this comparison was averaged over all 
10,000 simulated sequences.
Table \ref{randmoness_table} summarizes significance level of rejection for the null 
hypothesis, which assumes that the two distributions are drawn 
from the same underlying distribution (or the observed nulls 
are drawn from a random distribution). Apart from PSRs B0818$-$13 
and B2021+51 (where the significance is marginally lower - 
$>$ 82\%), the null hypothesis is rejected at high 
significance for the rest of the pulsars.

A stronger test is Wald$-$Wolfowitz statistical runs$-$test 
(Wald \& Wolfowitz 1940). A dichotomous data set, such as 
the nulling pattern, can be represented by a series of length 
n consisting of n$_1$ 1s (i.e. burst pulses) and n$_2$ 0s (i.e. null pulses), 
with each contiguous series of 1 or 0 defined as a run, r 
(i.e. number of runs given in Column 9 of Table \ref{Results}). 
In order to quantify the degree to which the runs are likely to 
represent a non-random sequence, we calculated Z, defined as  

\begin{equation}
Z = \frac{r - E(R)}{\sqrt{Var(R)}}
\label{Zeq}
\end{equation}

where, the mean of the random variable, R, is given by

\begin{equation}
E(R) = 1  + \frac{2 n_1 n_2}{n_1 + n_2}
\end{equation}

The variance of R is given by

\begin{equation}
Var(R) = \frac{2 n_1 n_2 (2 n_1 n_2 - n_1 - n_2)}{(n_1 + n_2)^2(n_1 + n_2 -1)}
\end{equation}

Sampling distribution of Z asymptotically tends to a standard 
normal distribution in case of large n with a zero mean and unity 
standard deviation. Therefore, Z will be close to 
zero for a random sequence and the value of Z, derived from R, can be used 
to test the hypothesis that the given sequence is random in a 
distribution free manner. Note that a sequence judged 
random by the runs test indicates that each observation in 
a sequence of binary events is independent of its 
predecessor.

This statistic was calculated for 8 pulsars for which NLH and BLH 
are presented in Figure \ref{nbhistall}. The results are given 
in Table \ref{randmoness_table}. The observed values of Z for all 
8 pulsars (except for PSR B0835$-$41) were large and hence  
the null hypothesis is rejected with more than 95 \% significance. 
Even for PSR B0835$-$41, the rejection significance is 92 \%, 
although this is not as significant as the other pulsars. 
Interestingly, Z is negative for all 8 pulsars suggesting that 
the null and burst pulses tend to occur in groups. 

The preceding two tests confirm that a null (or a burst), 
in the null-burst sequence for the 8 pulsars studied in this 
paper, is not independent of the state of the pulse preceding 
it. In other words, individual nulls (bursts) are correlated across 
several periods.  
However, these tests place no constraint on the randomness 
of the duration of nulls (bursts). A visual examination of 
data suggests that the  interval 
between two {\it transition events}, defined as a 
transition from a null to burst and vice-verse, 
does not depend on the duration of previous nulls or 
bursts and appears to be 
randomly distributed.
%{\bf Figure \ref{figdurseqnull} and \ref{figdurseqburst} 
% show the sequence of nulls and bursts respectively 
% for PSR B2111+46. The null and burst durations appear 
% to be organised randomly in the overall single pulse 
% sequence.} 
%follows 
%the distribution characterizing a 
%Poisson point process.} 

This random behaviour of the null (burst) duration 
is also supported by the following arguments. 
When the complete pulse sequence is divided into 
several subintervals, consisting of equal number of periods 
(typically 200 pulses), the count of the number of such 
transitions (events) is distributed as a Poisson distribution 
for all pulsars in our sample.  
Likewise, the interval between two transitions is distributed 
as an exponential distribution as is  evident in the NLH and BLH 
in Figure \ref{nbhistall}. Lastly, featureless spectra 
are obtained from the sequence of null(burst) durations 
indicating no correlations between these durations. 
Thus, it appears that the duration of nulls and bursts can 
be considered as a random variable, at least over the time 
scale spanned by our data for 
our sample of pulsars. 

In summary, the Wald$-$Wolfowitz runs tests imply non-randomness 
(i.e. correlation in the one-zero sequence, derived from the pulse 
sequence, across periods) in nulling in the sense that the absence (or presence) 
of emission in a given pulse is not independent of the state of 
the pulse preceding it, 
hinting a memory of the previous state.
However, the duration of the null and burst states
and the  time instants of these transitions appear to be random.
Hence, these pulsars produce nulls and bursts 
with unpredictable durations.

\section{Expected time scale for nulls and bursts}
\label{sec5a}

The nature of random variable, characterizing the null 
(burst) duration, is investigated further in this section, 
primarily to obtain the expected time scale for nulls (bursts). 
NLH and BLH in Figure \ref{nbhistall} suggests 
that the null and burst durations are distributed 
as an exponential distribution, which characterizes a stochastic 
Poisson point process. The CDF, F(x), of a Poisson point 
process is given by (Papoulis 1991)

\begin{equation}
F(x)  = 1  - \exp{(-x/\tau)}
\label{poiscdf}
\end{equation}

where, $\tau$ represents a characteristic 
time-scale of the stochastic process. A least square  
fit to this simple model  provides the 
characteristic null and burst 
time-scales ($\tau_n$ and $\tau_b$ respectively;
Equation \ref{poiscdf}).

\begin{figure}
 \centering
  \psfig{figure=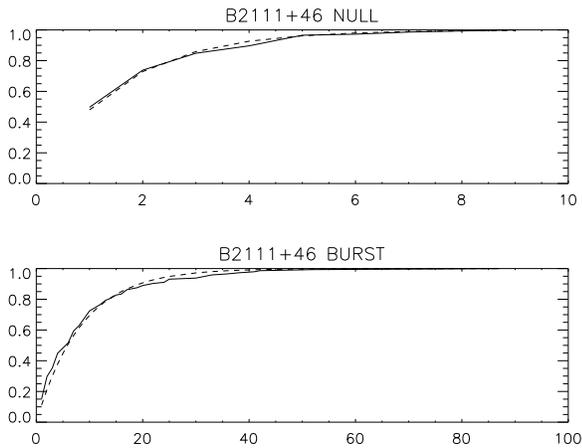,width=2.5in,angle=90}
\caption{CDF of null length (top plot) and burst length (bottom plot) 
distributions for PSR B2111+46 
(solid line) is shown alongwith the best fit Poisson point process 
model (dashed line), given by Equation \ref{poiscdf} }
\label{cdfall}
\end{figure}

Figure \ref{cdfall} shows 
the CDF (solid line) corresponding to NLH and BLH for PSR B2111+46 
alongwith a least square fit to the expected Poisson 
point process CDF given in Equation \ref{poiscdf}. These fits 
suggest that the interval between one transition from null 
to burst state (and vice-verse) to another transition 
from burst to null state appears to be modeled well by a 
Poisson point process\footnote{The model given in 
Equation \ref{poiscdf} need not be unique and other models 
may fit the CDFs equally well (See Vivekanand 1995). However, 
we use this model (a) as this is the simplest model suggested 
by our data, and (b) we did not have sufficient data 
to try more complicated models.} for this pulsar.

The characteristic null and burst 
time-scales ($\tau_n$ and $\tau_b$ respectively; 
Equation \ref{poiscdf}) and the uncertainties 
on these parameters, obtained from these fits,  
are listed in Table \ref{tabtau}. No fits were carried 
out for the CDF of nulls for PSRs B0835$-$41 and B2021+51 
as only two points  were available for the fit.
The fitted model was checked by carrying out a  two-sample KS-test 
in the following manner. 
First a pulse sequence, consisting 
of a million pulses, was simulated using the parameter 
$\tau$ obtained in these fits. Then, the NLH and BLH 
were obtained for this simulated pulse sequence, 
which provides much larger sample of nulls and bursts 
than the observed sequence. A two-sample KS-test was 
carried out on the NLH and BLH,  obtained from the observed 
sequence and the simulated pulse sequence. The significance 
level of rejection for the null hypothesis, which assumes, 
in this case, that the two distributions are different,   
is given in 
Column 4 and 6 of Table \ref{tabtau} for the null and the burst durations 
respectively. The null hypothesis is rejected with high 
significance for null duration in all pulsars (six), for 
which the fit was carried out. Apart from PSRs B2034+19  
and B2319+70\footnote{The lower significance of rejection 
of null hypothesis for the burst duration in these two 
pulsars may be due to the use of the simple model given by 
Equation \ref{poiscdf}.}, the null hypothesis is rejected with high 
significance for the burst duration for the other six pulsars.

\begin{center}
\begin{table}
% \scriptsize
\centering
\caption{KS statistic from comparison of null length distributions for 
four pulsars with similar NF of 1\%. 
The number in the parentheses beside the pulsar names are the number 
of null sequences used for comparison.
The number given in the table is the significance of 
rejecting null hypothesis, which assumes that the samples 
are drawn from the same distribution. The respective 
D value is given in the parentheses besides the significance of 
the test.}
\label{K-S comparison}
\begin{tabular}{l|c|c|c}
\hline
		& B0818$-$13(57)   & B0835$-$41(74)  & B2021+51(12)     \\ 
\hline
B0809+74(123)  &     (0.48) 99.9   &   (0.44) 99.9     &   (0.52) 97.9  \\ 
% \hline
B0818$-$13(57) &                   &   (0.2) 88.5      &   (0.46) 98.0  \\ 
% \hline
B0835$-$41(74) &                   &                  &   (0.48) 99.1   \\ 
\hline
\end{tabular}
\normalsize
\end{table}
\end{center}
\begin{center}
\begin{table}
% \scriptsize
\centering
\caption{Summary of the  two randomness tests, described in the text, for eight pulsars. The null hypothesis assumes that 
 every null is characterized by 
an iid random variable, for which NF represents the proportion 
statistics. Rejection of the above hypothesis with the rejection  significance from KS statistic and 
 runs test statistic are given in the Column 3 and 5 respectively. NF (reproduced from Table \ref{Results}) are given for comparison} 
\begin{tabular}[ht]{l|c|c|c|c}
\hline
PSRs       &  NF(\%)    &  KS test & Z &  Runs test   \\
\hline
B0809+74   &  1.42      & 99.9 & -38.29 &   99.9   \\
B0818$-$13 &  1.01      & 85.4 & -10.52 &   99.9   \\
B0835$-$41 &  1.7       & 98.7 & -2.77  &   92.1   \\
B1112+50   &  64      & 99.9 & -22.37 &   99.9   \\
B2021+51   &  1.4       & 82.2 & -15.72 &   99.9   \\
B2034+19   &  $\geq$26  & 99.9 & -12.53 &   99.9   \\
B2111+46   &  21      & 99.9 & -17.50 &   99.9   \\
B2319+60   &  29        & 99.9 & -43.10 &   99.9   \\
\hline
\label{randmoness_table}
\end{tabular}
\normalsize
\end{table}
\end{center}

%%%%%%%%%%%%%%%%%%%%%%%%%%%%%%%%%%%%%%%%%%%%%%%%%%%%%%%%%%%%%%%%%%%%%%%%%%%%%%%%
% Fitted CDF figure
%%%%%%%%%%%%%%%%%%%%%%%%%%%%%%%%%%%%%%%%%%%%%%%%%%%%%%%%%%%%%%%%%%%%%%%%%%%%%%%%

%%%%%%%%%%%%%%%%%%%%%%%%%%%%%%%%%%%%%%%%%%%%%%%%%%%%%%%%%%%%%%%%%%%%%%%%%%%%%%%%

%$$$$$$$$$$$$$$$$$$$$$$$$$$$$$$$$$$$$$$$$$$$$$$$$$$$$$$$$$$$$$$$$$$$$$$$$$$$$$
\begin{center}
\begin{table}
% \scriptsize
\centering
\caption{The characteristic null ($\tau_n$) and burst ($\tau_b$) timescale 
for the eight pulsars in Figure \ref{nbhistall} obtained from a least squares fit to 
the CDF of these pulsars to a Poisson point process. The time-scale 
has been expressed in seconds (i.e. column 3 and 5) after multiplying the fitted parameter 
$\tau$ in Equation \ref{poiscdf} with the period of the pulsar. The numbers 
in the parentheses are the corresponding errors on the time scales, 
obtained from the least squares fits. 
The Kolmogorov-Smirnov probability of rejection for the null hypothesis, 
which assumes that the two distributions are different, are given in 
column 4 and 6 for the null and burst durations respectively.} 
\begin{tabular}[ht]{l|c|c|c|c|c|}
\hline
PSRs       &  Period   &   $\tau_n$   &  KS-prob & $\tau_b$ & KS-prob  \\
           &   (s)     &    (s)       &    \%    &   (s)    &   \%     \\
\hline
B0809+74   & 1.29  &    1.9 (0.3)    & 100   & 176 (9)   & 98    \\
B0818$-$13 & 1.24  &    0.7 (0.3)    & 100   & 84 (9)    & 78    \\
B0835$-$41 & 0.75  &    -            & -     & 44 (4.5)  & 74    \\
B1112+50   & 1.66  &    4.8 (0.1)    & 100   & 4.3 (0.8) & 88    \\
B2021+51   & 0.53  &    -            & -     & 21 (10)   & 98    \\
B2034+19   & 2.07  &  2.6 (0.2) & 99    & 11 (2)    & 22    \\
B2111+46   & 1.02  &   1.6 (0.1)     & 99    & 8.7 (0.5) & 99    \\
B2319+60   & 2.26  &   11 (1)        & 94    & 23 (2)    & 33    \\
\hline
\label{tabtau}
\end{tabular}
\normalsize
\end{table}
\end{center}

%$$$$$$$$$$$$$$$$$$$$$$$$$$$$$$$$$$$$$$$$$$$$$$$$$$$$$$$$$$$$$$$$$$$$$$$$$$$$$
%\begin{center}
%\begin{table}
% \scriptsize
%\centering
%\caption{The characteristic null ($\tau_n$) and burst ($\tau_b$) timescale 
%for the eight pulsars in Figure \ref{nbhistall} obtained from a {\bf least squares} fit to 
%the CDF of these pulsars to a Poisson point process. The time-scale 
%has been expressed in seconds after multiplying the fitted parameter 
%$\tau$ in Equation \ref{poiscdf} with the period of the pulsar. The numbers 
%in the parentheses are the corresponding errors on the time scales, 
%{\bf obtained from the least squares fits} .}
%\begin{tabular}[ht]{l|c|c|c|c|}
%\hline
%PSRs       &  Period   & NF   &  $\tau_n$ & $\tau_b$    \\
%           &   (s)     & (\%) &   (s)     &   (s)       \\
%\hline
%B0809+74   & 1.292241  & 1.42 &   1.67 (0.04)     &   99 (4)  \\
%B0818$-$13 & 1.238130  & 1.01 &   0.56 (0.03)     &   85 (4)  \\
%B0835$-$41 & 0.751624  & 1.7  &   0.21 (0.001)    &   44 (1)  \\
%B1112+50   & 1.656440  & 64   &   4.79 (0.04)     &   2.95 (0.11)   \\
%B2021+51   & 0.529197  & 1.4  &   0.49 (0.15)     &   21 (4)  \\
%B2034+19   & 2.074377  & $\geq$ 26 & 2.50 (0.05)  &   8.3 (0.6)   \\
%B2111+46   & 1.014685  & 21   &  1.55 (0.03)      &   8.2 (0.2)   \\
%B2319+60   & 2.256488  & 29   &  10.05 (0.43)     &   21.2 (0.8)   \\
%\hline
%\label{tabtau}
%\end{tabular}
%\normalsize
%\end{table}
%\end{center}

%\section{Conclusions and Discussion}
%\label{sec6}

\section{Discussion}
\label{sec6}
The estimates for the factor, $\eta$, by which the pulsed emission 
reduces during the nulls for 11 pulsars were presented for the 
first time in this paper (Table 1). Although the physical process,  
which causes nulling, is not yet understood, it could be due to a 
loss of coherence in the plasma generating the radio emission or 
due to geometric reasons. In the former case, $\eta$ provides 
a constraint on the process responsible for this loss of coherence. 
Our estimates provide lower limits for different pulsars as 
this estimate is limited by the available S/N. Nevertheless, 
reduction by two orders of magnitude is seen in at least two 
pulsars. If nulling is caused by a shift in the radio beam due to 
global changes in magnetosphere, $\eta$ provides a constraint 
on the low level emission and will depend on the orientation 
to the line of sight and the morphology of the beam during the 
null. 

Our results confirm that NF probably does not capture 
the full detail of the nulling behaviour of a pulsar. 
We find that the pattern of nulling can be quite different 
for classical nulling pulsars with similar NFs. Estimates for 
typical timescales, $\tau_n$, for 6 pulsars in our sample were 
obtained for the first time. For 2 of these with a NF of about 
1 percent, $\tau_n$ varies by a factor of three. In particular, 
the typical nulling timescale for PSR B1112+50  
(NF $\sim$ 65\%) is about 2 s, more than 6 orders of magnitude 
less than that for the intermittent pulsar 
PSR B1931+24 (NF $\sim$ 75\%). In the Ruderman and Sutherland (1975) model, the 
pulsar emission is related to relativistic pair plasma generated due 
to high accelerating electric potential in the polar cap. 
Changes in this relativistic plasma flow (Filippenko \& Radhakrishna 1982; 
Lyne et al. 2010), probably caused by changes in the polar cap 
potential, have been proposed as the underlying cause 
for a cessation of emission during a null. The typical nulling 
timescale, $\tau_n$ (and burst timescale $\tau_b$) provides a 
characteristic duration for such a quasi-stable state, which is 
similar to a profile mode-change. An interesting possibility 
may be to relate this to the polar cap potential. In any case, 
any plausible model for 
nulling needs to account for the range of null durations for 
pulsars in our sample and relate it to a physical parameter 
and/or magnetospheric conditions in the pulsar magnetosphere.

We have extended the Wald-Wolfowitz runs test for randomness 
to 8 more pulsars. Results for 15 pulsars were  published 
in previous studies (Redman and Rankin 2009; Kloumann and Rankin 2010). 
Our sample  has no overlap with  the sample presented 
by these authors. Like these authors, we find that this test indicates that 
occurrence of nulling, when individual pulses are considered, is 
non-random or exhibits correlation across periods. 
Unlike these authors, all 8 pulsars in our sample show such a behaviour. 
This correlation groups pulses in null and burst states, which was also 
noted by Redman and Rankin (2009). 
However, the duration of the null and burst states seems to be modeled 
by a stochastic Poisson point 
process suggesting that these transitions occur at random. 
Thus, the underlying physical process for nulls in the 8 pulsars,  
studied in this paper, appears to be random in nature producing 
nulls and bursts with unpredictable durations.

Lastly, our estimates of NF in Table 1 are consistent with those published earlier 
for PSRs B0809+74, B0818$-$13, B0835$-$41, B1112+50 and 
B2319+60 (Lyne \& Ashworth 1986; Biggs 1992; Ritchings 1976), 
which indicates that NF are consistent over a 
time scale of about 30 years. 

\section{Conclusions}
\label{sec7}
The nulling behaviour of 15 pulsars,  out of which 5 were 
PKSMB pulsars with no previously reported nulling behavior, 
is presented in this paper with estimates of their NFs. For 
four of these 15 pulsars, only an upper/lower limit was previously 
reported. The estimates of reduction in the pulsed emission 
is also presented for the first time in 11 pulsars. NF value for 
individual profile component is also presented for two pulsars in the 
sample, namely PSRs B2111+46 and B2020+28. Possible 
mode changing behaviour is suggested by these observations 
for PSR J1725$-$4043, but this needs to be confirmed  with 
more sensitive observations. 
An interesting quasi-periodic nulling behaviour for 
PSR J1738$-$2330 is also reported. 
We find that the nulling patterns differ between  
PSRs B0809+74, B0818$-$13, B0835$-$41 and B2021+51, 
even though they have similar NF of around 1\%. 

The null and burst pulses in 8 pulsars in our sample appear 
to be grouped and seem to occur in a correlated way, when 
individual periods are considered. However, 
the interval between  transitions from the null to the  burst states
(and vice-verse) appears   
to represent a Poisson point process. The typical null 
and burst timescales for these pulsars have been obtained 
for the first time to the best of our knowledge.

Pulsar nulling remains an open question even after 
40 years since it was first reported. Recent studies 
suggests that both nulling and  profile mode changes probably 
represent a global reorganization of pulsar magnetosphere 
probably accompanied by changes in the 
spin-down rate of these pulsars (Kramer et al. 2006; Lyne et al. 2010). 
Interestingly, such quasi-stable states of magnetosphere 
have recently been proposed, based on MHD calculations, 
to explain the release of magnetic energy implied by high energy bursts 
in soft-gamma ray repeaters (Contopoulos, Kazanas \& 
Fendt 1999; Contopoulos 2005; Timokhin 2010). 
Global changes in magnetospheric state is likely to be 
manifested in changes in radio emission regardless of the 
frequency of observations. Future simultaneous multifrequency observations 
of pulsars with nulling will be useful to study these 
changes and constrain such magnetospheric models.

\section*{Acknowledgments}
We would like to thank staff 
of GMRT and NCRA for providing valuable support in carrying out this project. 
Authors also thank Avinash Deshpande and D. J. Saikia 
for valuable comments and discussions. 
VG would like to thank Dipanjan Mitra for valuable discussion regarding 
nulling in general. We thank Mihir Arjunwadkar for useful discussions 
on the statistical techniques and random processes. 
Authors also thank the anonymous referee 
for useful suggestions and criticism, which helped in 
improving the manuscript.

%\acknowledgements %%% Text of acknowledgements runs on after this command.
%The Giant Meterwave Radio Telescope is a part of project by National Center for Radio Astrophysics which is funded by Tata Institute of Fundamental Research and Department of Atomic Energy.
%%% THE BIBLIOGRAPHY
%%%
%%% CONSULT SECTION 3 OF "INSTRUCTIONS FOR AUTHORS" FOR HOW TO USE NATBIB.
%%% AUTHORS ARE ENCOURAGED TO USE EITHER THE "THEBIBLIOGRAPY" ENVIRONMENT
%%% BY UNCOMMENTING (DELETING THE "%" SYMBOL) THE COMMANDS BELOW, OR BY
%%% USING THE BIBTEX ENVIRONMENT. TO FIND OUT WHICH IS APPLICABLE TO YOUR
%%% CONTRIBUTION, CONSULT THE VOLUME EDITORS FOR YOUR PROCEEDINGS.
%%%

\end{document}